\documentclass[aip,amsfonts,amsmath,amssymb,reprint]{revtex4-2}
\usepackage{graphicx}
\usepackage{dcolumn}
\usepackage{bm}

\usepackage{xcolor}
\usepackage[english]{babel}
\usepackage[utf8]{inputenc}
\usepackage[T1]{fontenc}
\usepackage{mathptmx}
\usepackage{tabularx,booktabs}
\usepackage{soul}
\usepackage{color}

\usepackage{soul, xcolor}
\setstcolor{red}
\begin{document}

\title{Topological analysis of the latent geometry of a complex network}
\author{Bukyoung Jhun}
\email{jhunbk@snu.ac.kr}
\affiliation{CCSS, CTP and Department of Physics and Astronomy, Seoul National University, Seoul 08826, Korea}
\affiliation{Department of Physics, The University of Texas at Austin, Austin, TX 78712, USA}

\date{\today}

\begin{abstract}
Most real-world networks are embedded in latent geometries. If a node in a network is found in the vicinity of another node in the latent geometry, the two nodes have a disproportionately high probability of being connected by a link.
The latent geometry of a complex network is a central topic of research in network science, which has an expansive range of practical applications such as efficient navigation, missing link prediction, and brain mapping.
Despite the important role of topology in the structures and functions of complex systems, little to no study has been conducted to develop a method to estimate the general unknown latent geometry of complex networks.
Topological data analysis, which has attracted extensive attention in the research community owing to its convincing performance, can be directly implemented into complex networks; however, even a small fraction (0.1\%) of long-range links can completely erase the topological signature of the latent geometry.
Inspired by the fact that long-range links in a network have disproportionately high loads, we develop a set of methods that can analyze the latent geometry of a complex network: the modified persistent homology diagram and the map of the latent geometry.
These methods successfully reveal the topological properties of the synthetic and empirical networks used to validate the proposed methods. 
\end{abstract}

\maketitle

\begin{quotation}
If two people went to the same school or worked in the same company (i.e. close in a social landscape), they are more likely to be connected in the friendship network than people who did not.
Such a hypothetical landscape is the latent geometry of a network: if two nodes in a network are found close to each other in the latent geometry, they have a higher probability of being connected by a link.
The latent geometry of a complex network is a central topic of research in network science with extensive practical applications.
However, even if the two nodes are distant from each other in the latent geometry, there is a small probability that they are connected by a link.
Such long-range links complicate the discovery of the latent geometry of a network.
In this study, we developed a framework that can analyze the topology of the latent geometry of a complex network.
We tested the proposed methods and demonstrated that they successfully estimate the topological properties of synthetic and empirical networks.
\end{quotation}

\section{Introduction}

Many networks are embedded in latent geometries.
For instance, in the Watts–Strogatz small-world network~\cite{watts1998}, each node is embedded at a point on a ring, and if two nodes are close to each other on the ring, there is a disproportionately high probability that they are connected by a link. 
Such hypothetical space where each node is embedded is called the latent geometry, and a network with a latent geometry is called a noisy geometric network~\cite{taylor2015}.
If two nodes are close in the latent geometry, they have a high probability of being connected by a link.
Short-range links in the latent geometry are called geometric edges, and long-range links are called nongeometric edges.
Examples of noisy geometric networks also include the variants of the small-world network~\cite{kleinberg2000,newman2000} and some scale-free network models~\cite{rozenfeld2002,kim2004a}.
Another example is the popularity-similarity optimization model~\cite{krioukov2010,boguna2010,papadopoulos2012,jhun2021}, in which the latent geometry is a hyperbolic space.

The latent geometry of a complex network is a central topic of research in network science.
Latent geometry has been used to efficiently navigate networks~\cite{kleinberg2000,boguna2010}, predict missing links~\cite{clauset2008,liben-nowell2007}, and map the brain~\cite{cacciola2017}.
Certain epidemic phenomena spread along the latent geometry~\cite{taylor2015}.
Embedding a given network into a hyperbolic geometric space has been extensively studied ~\cite{muscoloni2017,alanis-lobato2016,krioukov2010} with a wide variety of applications.
Other topological studies of networks have focused on the number of small-scale homologies, which is typically proportional to the system size~\cite{horak2009,petri2013,bobrowski2020,carstens2013,lee2021,oh2021}.
However, the estimation of the latent geometry of a complex network has attracted little attention~\cite{taylor2015}.
In this study, we use three noisy geometric network models with various latent geometries along with real-world networks to test whether our methods estimate the correct properties of a given latent geometry.

Topological data analysis (TDA) has attracted extensive attention in recent years~\cite{carlsson2009,wasserman2018,sizemore2019}.
In the TDA, it is assumed that there is a latent geometry in the configuration space where the data points lie, and the points are stochastically generated on the latent geometry.
The objective of TDA is to estimate the topology of latent geometry and its properties.
One of the most prominent methods in TDA is persistent homology~\cite{carlsson2005,zomorodian2004}.
Persistent homology estimates the topological features of the latent geometry such as the number of connected components, holes, and cavities.
Persistent homology has been used as a topological tool to reveal how cognitive circuits are programmed~\cite{chaudhuri2019}, analyze the relationship between the topology and function of proteins ~\cite{xia2014}, identify brain disease~\cite{lee2019}, analyze the effect of psychedelics on the functional pattern of brain activity~\cite{petri2014}, identify topological phase transitions~\cite{park2021}, classify nanoporous materials~\cite{lee2017}, and analyze structures of social collaborations~\cite{carstens2013,gao2019}.

Another notable method is the mapper~\cite{singh2007}.
The method summarizes the given set of high-dimensional points into a mapper graph.
Data points are assigned to nodes in the resulting mapper graph, which represents the simplified latent geometry of the data.
The mapper has been used in a wide range of disciplines to discover a new subgroup of breast cancer~\cite{nicolau2011}, identify attention-deficit/hyperactivity disorder~\cite{kyeong2015} and neurological damage~\cite{nielson2015}, reveal the dynamic organization of the brain~\cite{saggar2018}, distinguish resilience to infection in the human population~\cite{torres2016}, predict manufacturing productivity~\cite{guo2016}, and mine data in social media~\cite{almgren2017}.
These methods of TDA, persistent homology and mapper, can be implemented into networks; however, as we will demonstrate, the presence of only 0.1\% long-range links is sufficient to completely erase any topological signature of the network.

Herein, we propose a set of systematic methods that can estimate the latent geometry of a given network: the modified persistent homology diagram and the map of the latent geometry.
To test the proposed methods, we applied them to various synthetic noisy geometric network models with various latent geometries.
The proposed methods successfully estimate the latent geometries of the models.
We also applied these methods to an empirical network, the co-authorship structure of researchers in network science.

The remainder of this article is organized as follows: In Sec.~\ref{sec:noisy_geometric}, we explain the concept of a noisy geometric network and introduce some examples that are used to verify the proposed methods.
We introduce the modified persistent homology diagram in Sec.~\ref{sec:persistent_homology} and demonstrate that we can recover the latent geometry by removing the links with high loads.
In Sec.~\ref{sec:map}, the map of the latent geometry is presented.
To validate the proposed framework for the topological analysis of complex networks, we test the methods on synthetic and empirical networks in Sec.~\ref{sec:numerical_result}.
The summary and final remarks are presented in Sec.~\ref{sec:conclusion}.

\section{Noisy geometric networks}
\label{sec:noisy_geometric}

\begin{figure*}
	\centering
	\includegraphics{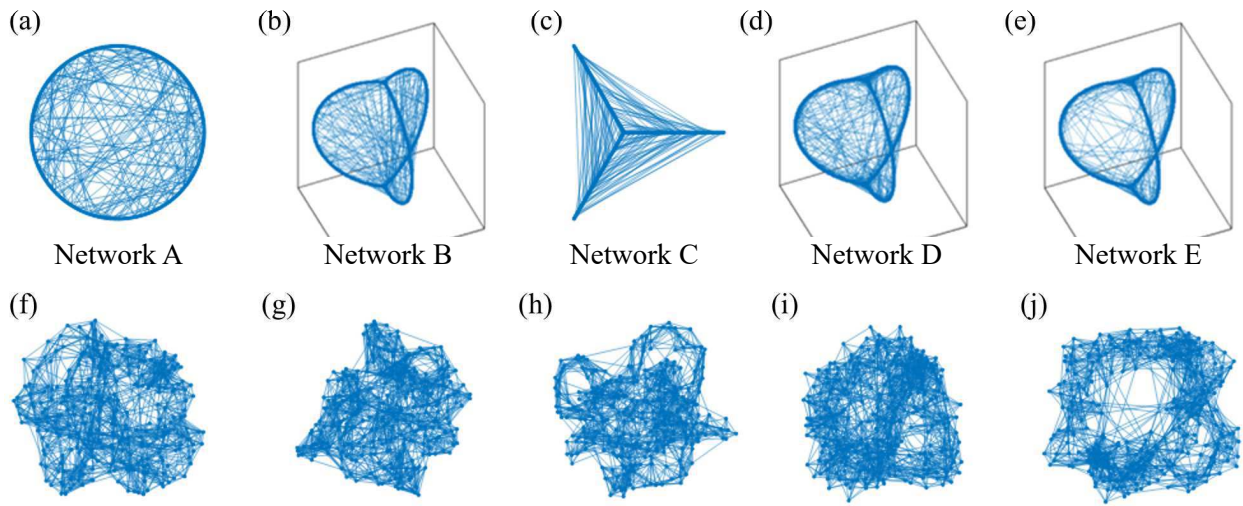}
	\caption{
		Various types of noisy geometric networks with various latent geometries used to validate the proposed methods.
		Network A: Watts–Strogatz small-world network with a latent ring structure laid out (a) on the latent geometry and (f) by the force layout algorithm~\cite{fruchterman1991}.
		Network B: Watts–Strogatz small-world network with a double-ring latent geometry laid out (b) on the latent geometry and (g) by the force layout algorithm.
		Network C: Watts–Strogatz small-world network with a triple-line latent geometry laid out (c) on the latent geometry and (h) by the force layout algorithm.
		Network D: Multilayer network of an exponentially connected network and an Erd{\H o}s--R\'enyi network with a double-ring latent geometry laid out (d) on the latent geometry and (i) by the force layout algorithm.
		Network E: Power-law connected network with a double-ring latent geometry laid out (e) on the latent geometry and (j) by the force layout algorithm.
		The number of nodes in each network is 1200, and the mean degree is 12.
		The fraction of long-range links is $P=0.1$ in small-world networks and the multilayer network.
		The power of connectivity is $\sigma=2$ in the power-law connected network.
		When a force layout algorithm is used to layout the network instead of using the information of the latent geometry, the latent geometry is completely lost, as illustrated in (f--j).
	}
	\label{fig2}
\end{figure*}

In this study, we used three noisy geometric network models with various latent geometries along with a real-world network to test whether the methods estimate the correct properties of a given latent geometry.
The Watts–Strogatz small-world network~\cite{watts1998} was proposed as a model that manifests a high clustering coefficient and short average path lengths and has served as a classic network model in the research community along with its variants~\cite{kleinberg2000,newman2000}.
The original Watts–Strogatz model had a ring as its latent geometry.
The latent geometry serves as a metric space, where the distance between two nodes is measured along the structure.
The connection probability of two nodes in a Watts–Strogatz network is solely determined by whether the distance between the two nodes in the latent geometry is smaller or larger than $K/2$.
If the parameters of the network are $N$, $K$, and $P$, two points with a distance of less than $K/2$ between them are directly connected by a link with a probability of $1-P+KP^2/(N-1)\simeq 1-P$.
If the distance is larger than $K/2$, the probability is $KP/(N-1)$, which approaches zero as $N$ approaches infinity while $K$ and $P$ are fixed.
A small fraction $P$ of long-range links gives the small-world property to the network.
In this study, we constructed Watts–Strogatz small-world networks with the latent geometries of a ring (Network A), double-ring (Network B), and triple-line (Network C) to validate the proposed methods [Fig.~\ref{fig2}(a--c), respectively].

Short-range links are highly regular in the Watts–Strogatz model.
To show that the proposed findings do not rely on this regularity, we construct a model with stochastically generated short-range links: the \textit{exponentially connected network}.
An exponentially connected network is constructed on a latent geometry, and each node is assigned to a point in the structure.
The probability that two nodes $i$ and $j$ are connected by a link is given by an exponential function of the distance between them,
\begin{equation}
P_{ij} = e^{-d_{ij}/\xi} \,,
\end{equation} 
where $d_{ij}$ is the distance between the two nodes in the latent geometry and $\xi$ is the correlation length that characterizes the network.
An exponentially connected network with a one-dimensional latent geometry has a mean degree $\left<k\right> = 2\sum_{i=1}^\infty e^{-i/\xi} = 2/(e^{1/a}-1)$.
We then construct a multilayer network with one layer consisting of an exponentially connected network with correlation length $\xi$ and another consisting of an Erd{\H o}s--R\'enyi network with mean degree $\left<k_{LR}\right>$.
The links in the exponentially connected network serve as short-range links and the links in the Erd{\H o}s--R\'enyi network serve as long-range links.
The fraction of long-range links in the multilayer network is $\left<k_{LR}\right>/(\left<k_{LR}\right>+\left<k_{SR}\right>) = \left<k_{LR}\right>/(2/(e^{1/a}-1)+\left<k_{LR}\right>)$.

Long-range and short-range links are generated by different processes in the two aforementioned noisy geometric network models.
In the \textit{power-law connected network}, long-range and short-range links are created using the same process.
Similar to the exponentially connected network, each node of the network is assigned a position on the latent geometry.
Two points $i$ and $j$ in a power-law connected network are connected with the probability
\begin{equation}
P_{ij} = \left.
\begin{cases}
\alpha d_{ij}^{-\sigma} &  \left(d_{ij} \geq \alpha^{1/\sigma}\right) \\
1 &  \left(d_{ij} < \alpha^{1/\sigma}\right) \\
\end{cases}
\right. \,,
\end{equation}
where $d_{ij}$ is the distance between nodes $i$ and $j$ in the latent geometry, and $\sigma$ is the power that controls the decay of probability over distance.
The mean degree of a power-law connected network with a one-dimensional latent geometry is $\left<k\right> = 2\alpha\left(\zeta(\sigma)-\sum_{n=1}^{[\alpha^{1/\sigma}]+1}n^{-\sigma}\right) + 2\alpha^{1/\sigma}+2$.
In this study, we constructed a multilayer network of an exponentially connected network and an Erd{\H o}s--R\'enyi network with a double-ring latent geometry (Network D), and a power-law connected network with a double-ring latent geometry (Network E) to validate the proposed methods [Fig.~\ref{fig2}(d, e), respectively].

\section{Modified persistent homology diagram}
\label{sec:persistent_homology}

\begin{figure*}
	\centering
	\includegraphics{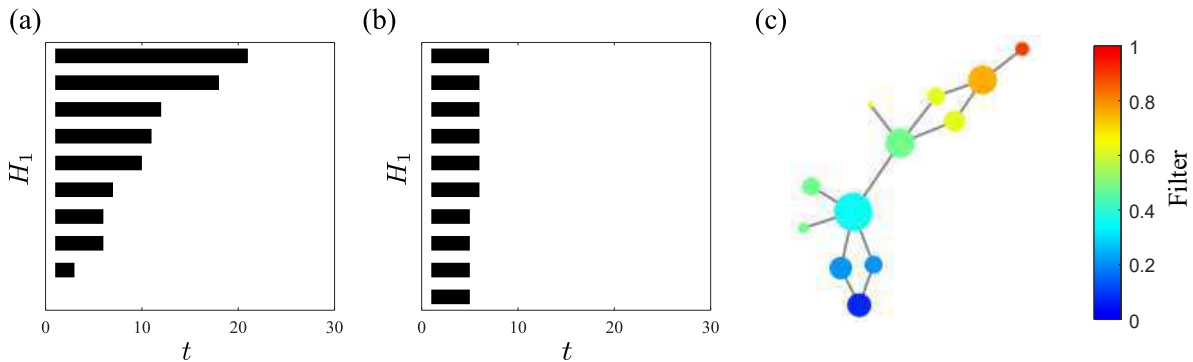}
	\caption{
		One-dimensional persistent homology $H_1$ of the Watts–Strogatz small-world network with $N=1200$, $K=12$, and (a) $P=0.001$ and (b) $P=0.01$.
		The latent geometry is a ring.
		Ten most persistent (longest) barcodes are illustrated.	
		The homology of the latent geometry is $H_1=1$; hence a single distinctively long bar should appear in the barcode; however, the method fails to reveal the homology.
		(c) The mapper graph of the Watts–Strogatz small-world network with $N=1200$, $K=12$, and $P=0.001$.
		The size of a vertex in the map is proportional to the number of network nodes in the cluster; the color encodes the value of the filter function, with red indicating high values and blue indicating low values.
		The mapper fails to reveal the latent geometry, which is a ring.
	}
	\label{fig1}
\end{figure*}

\begin{figure*}
	\centering
	\includegraphics{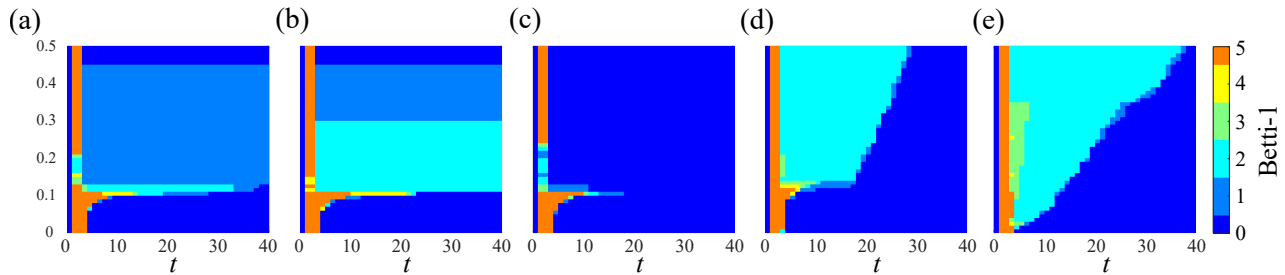}
	\caption{
		Modified persistent homology diagrams of the Watts–Strogatz small-world networks with latent geometries of (a) a ring (Network A), (b) a double-ring (Network B), (c) a triple-line (Network C), (d) a multilayer network of an exponentially connected network and an Erd{\H o}s--R\'enyi network with a double-ring latent geometry (Network D), and (e) a power-law connected network with a double-ring latent geometry (Network E).
		The diagram in (a) shows a persistent region of homology $H_1=1$ and correctly estimates the homology of a ring.
		The diagram in (b, d, e) shows a persistent region of homology $H_1=2$, and the diagram in (c) shows no persistent region of nonzero homology $H_1$.
		The number of nodes is 1200, and the mean degree is 12 in each network.
		The fraction of long-range links is $P=0.1$ in small-world networks and the multilayer network.
		The power of connectivity is $\sigma=2$ in the power-law connected network.
	}
	\label{fig3}
\end{figure*}

TDA investigates the latent shape of data using the distance between each pair of points.
Then, a simplicial complex is constructed by connecting a set of points that are close to each other with simplices using certain criteria.
A simplicial complex is a class of hypergraphs that satisfies a certain property: if a hyperedge $E$ is in a simplicial complex $H$, any hyperedge $E^\prime\subset E$ is also in $H$.
Once a simplicial complex is constructed, we implement methods of algebraic topology to calculate the topological properties of the simplicial complex.
We then use the obtained results to estimate the properties of the latent geometry of the given data set.
The most common simplicial complex used in TDA is the Vietoris--Rips complex (often called the Rips complex).
For a radius $t$, a one-dimensional simplex (link) $S_1=\{i_0,i_i\}$ is included in the Rips complex if and only if the distance between nodes $i_0$ and $i_1$ is no greater than $t$: $d(i_0,i_1)\leq t$.
A higher $d$-dimensional simplex $s_d=\{i_0,i_1,\cdots,i_d\}$ is included in the Rips complex if and only if the distance between all pairs in the simplex is less than $t$: $\forall x,y \in s_d$, $d(x,y)\leq t$.
The Rips complex generated by this rule satisfies the properties of the simplicial complex: if a simplex $s_d$ is in a simplicial complex $S$, any simplex $s^\prime \subset s_d$ is also in $S$.

Persistent homology~\cite{carlsson2005,zomorodian2004}, one of the most commonly used tools in TDA, encodes the multiscale topological features of data.
For a standard $\epsilon$-ball persistent homology, a sequence of the Rips complex is constructed for a range of radii $t$.
As the radius $t$ increases, the data points are connected to form loops, voids, and higher-dimensional holes in the Rips complex.
As $t$ further increases, the loops, voids, and higher-dimensional holes in the Rips complex are eventually filled.
Zero-dimensional homology measures the number of connected components, one-dimensional homology measures the number of loops, and two-dimensional homology measures the number of voids in the simplicial complex.
Higher-dimensional homology measures the number of higher-dimensional holes in a simplicial complex but is not commonly used in applications because it requires excessive computation and little is known about the interpretations of higher-dimensional Betti numbers.

The radii where holes are formed and filled are tracked and summarized as a barcode diagram.
A hole that is formed at $t_1$ (birth radius) and filled at $t_2$ (death radius) is depicted by a bar from $t_1$ to $t_2$ in the diagram.
A large structure in the data points, which we regard as a significant feature of the data, is represented by long bars in the barcode.
Short bars are deemed as noise.
Persistent barcodes are stable under perturbations of data~\cite{cohen-steiner2007}.
If $n$ number of bars are distinctively longer than others in the $d$-dimensional persistent homology diagram, it suggests that for a wide range of $t$, there are $n$ $d$-dimensional holes in the Rips complex; therefore, we conclude that there are $n$ $d$-dimensional holes in the latent geometry of the data.

Only the distances between the pairs of points in the data are needed to implement persistent homology.
Therefore, we can apply this method to a complex network using the network distance, which is the length of the shortest path between two nodes.
However, when applied to a Watts–Strogatz small-world network with a ring latent geometry, it fails to estimate the latent geometry of the network [Fig.~\ref{fig1}(a, b)].
If the method successfully estimated the homology of the latent geometry, a single distinctively long bar would have appeared in the barcode.
The presence of only 0.1\% long-range links ($P=0.001$) is sufficient to completely erase the topological signature of the latent geometry.
Additionally, when a force layout algorithm~\cite{fruchterman1991} is used to layout the network without using the information of the latent geometry, the latent geometry is completely lost, as illustrated in Fig.~\ref{fig2}(f--j).

In a network, the birth radius of zero- and one-dimensional homology is always one.
For one-dimensional homology, it suffices to keep track of the homology at each radius, instead of keeping track of all the birth-death radii.
The direct implementation of persistent homology on the network counts the number of connected components, loops, and higher-dimensional holes.
Such loops and voids are formed stochastically, and their numbers are often proportional to the number of nodes in the network~\cite{horak2009,bobrowski2020}.
These microscopic properties of a network can be useful in some contexts but do not estimate the topological features of the latent geometry.

If there are no long-range links in the network, persistent homology can successfully estimate the homology of the latent geometry.
Therefore, if we identify and remove long-range links in a network, persistent homology can be implemented to estimate the number of loops and voids in the latent geometry.
One property that distinguishes long-range links from short-range links is the load~\cite{goh2001}, which is related to the number of shortest paths passing through a node or link because long-range links have significantly higher loads than those of short-range links.
Therefore, we distill the network by removing a certain fraction $\delta$ of the links with the highest loads.
Subnetworks consisting of high-load links have been extensively studied~\cite{goh2006,kim2004b,wu2006}, but subnetworks consisting of links with low loads have attracted little to no attention.
We varied the distillation rate $\delta$ along with the radius $t$ for a range of values.
Here, we focused on one-dimensional homology and studied the number of loops in the latent geometry, but the method can be generalized to study higher-dimensional homology.
If there are $n$ loops in the latent geometry, for a significant range of values of the distillation rate $\delta$, the one-dimensional homology is $n$ for a significant range of radius $t$.
Therefore, the one-dimensional homology should persistently appear as $n$ in a wide region on the $t$--$\delta$ plane (Fig.~\ref{fig3}).
This allows us to estimate the homology of the latent geometry of the network. 

To test the proposed method, we applied it to various types of synthetic noisy geometric networks described in Sec.~\ref{sec:noisy_geometric} and illustrated in Fig.~\ref{fig2} (Networks A--E).
The modified persistent homology diagrams illustrated in Fig.~\ref{fig3} clearly show a one-dimensional homology of 0 for the triple-line network, 1 for the ring network, and 2 for the double-ring networks.
The modified persistent homology for complex networks successfully estimates the homology of the latent geometries of synthetic noisy geometric networks.

\section{Map of the latent geometry}
\label{sec:map}

\begin{figure*}
	\centering
	\includegraphics{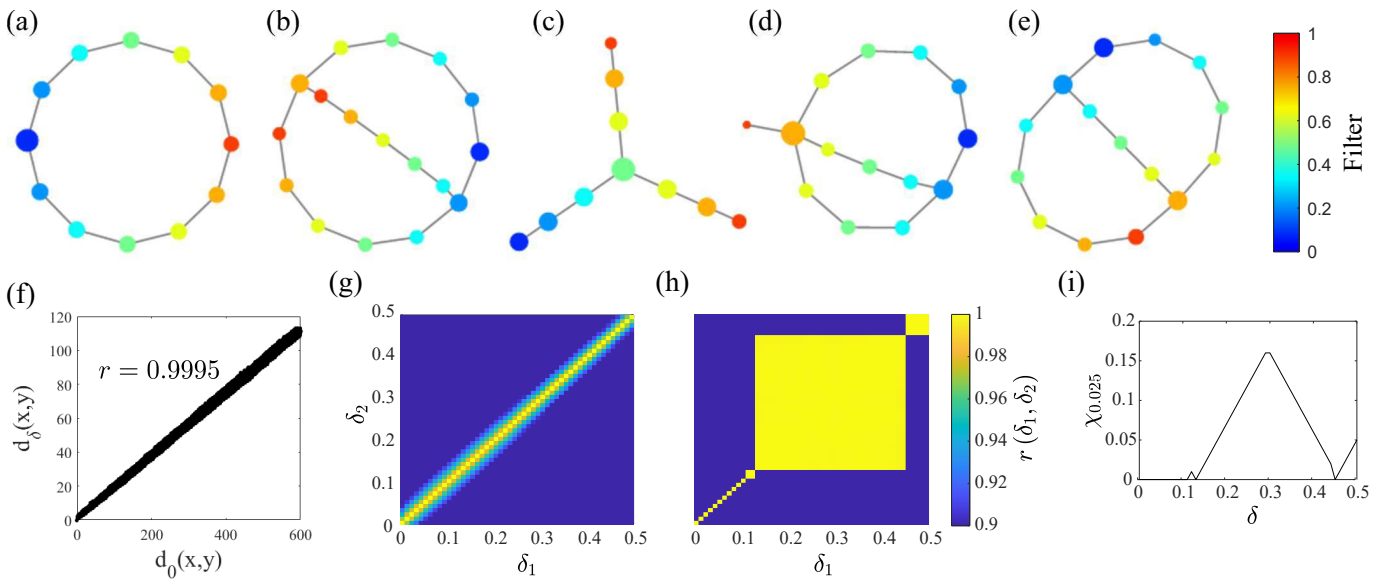}
	\caption{
		The map of the latent geometry of the Watts–Strogatz small-world network with (a) a ring latent geometry (Network A), (b) a double-ring latent geometry (Network B), (c) a triple-line latent geometry (Network C), (d) a multilayer network of an exponentially connected network and an Erd{\H o}s--R\'enyi network with a double-ring latent geometry (Network D), and (e) power-law connected network with a double-ring latent geometry (Network E).
		The size of a vertex in the map is proportional to the number of network nodes in the cluster, and the color encodes the value of the filter function, with red indicating high values and blue indicating low values.
		The number of nodes is 1200, and the mean degree is 12 in each network.
		The fraction of the long-range links is 0.1 in the small-world networks and the multilayer network.
		The power of connectivity is $\sigma=2$ in the power-law connected network.
		The map correctly represents the latent geometry in each case.
		(f) Scatterplot of the distances between pairs of nodes in the latent geometry and the distilled network.
		The Pearson correlation coefficient is 0.9995; hence, the distilled network successfully recovers the latent geometry.
		The Pearson correlation coefficient of distances on distilled networks $r(\delta_1,\delta_2)$ of (g) the Erd{\H o}s--R\'enyi network and (h) the Watts–Strogatz small-world network used in (a).
		A clear persistent interval appears in the small-world network but not in the Erd{\H o}s--R\'enyi network.
		(i) The persistence interval measured by $\chi_{0.025}$.
		A clear peak at $\delta=0.31$ appears.
	}
	\label{fig4}
\end{figure*}

Another prominent method in TDA is the mapper~\cite{singh2007}.
The purpose of the mapper is to simplify the latent geometry of a given set of data points as a mapper graph.
Unlike persistent homology, this requires the user to make a few choices of parameters.
The first thing that must be selected is the filter function.
The filter function is used to partition data points into groups.
A one-dimensional filter function $f$ assigns a real number to each data point $x \in X$, $f: X\rightarrow \mathbb{R}$.
A one-dimensional filter is often used to reveal a one-dimensional filamentary structure~\cite{reeb1946,shinagawa1991}, but the target space for the filter function can be selected as a higher-dimensional space, $f: X\rightarrow \mathbb{R}^d$.
Such a choice would reveal a higher-dimensional latent geometry.
We then divide the range of the function $I\subset \mathbb{R}$ of the given points into a set of smaller overlapping intervals $I_j$.
In this step, we select two parameters: the length of the smaller interval and the percentage overlap between successive intervals.
Then, for each interval $I_j\in S$ and $X_j=\{x|f(x)\in I_j\}$, we implement a clustering algorithm to find clusters $X_{jk}$ in each $X_j$.
In this step, we choose a parameter for the clustering algorithm of choice (for instance, the $k$ value in the $k$-nearest neighbor clustering or the number of clusters in agglomerative clustering).
The mapper does not impose any conditions on the clustering algorithm.
We represent each cluster $X_{jk}$ as a vertex in the mapper graph.
If there exists an overlapping point in two clusters (with distinct $j$), we connect the two clusters in the mapper graph.

The mapper can be implemented in a noisy geometric network using the network distance as the metric function.
The distance from a selected node is defined for each node that is in the same connected component as the selected node and can, therefore, be used as the filter function.
At each layer of the filter function $X_j$, we define each connected component in the subnetwork consisting of $X_j$ as $X_{jk}$.
We applied this algorithm to a Watts–Strogatz small-world network with $N=1200$, $P=0.001$, and $K=12$ [Fig.~\ref{fig1}(c)].
The mapper does not show a graph representing the latent geometry of the network, which is a ring.
Long-range links constituting 0.1\% of the total links completely erase the topological signature of the latent geometry.
Furthermore, network coarse-graining methods~\cite{gfeller2007,lee2010} do not show the latent structure in the presence of long-range links.

Similar to the case of the modified persistent homology diagram, we can remove a fraction of the links in the network with high loads and then apply the mapper algorithm or coarse-graining methods.
We distill the given network by removing a fraction $\delta$ of the links with the highest loads.
For an appropriate distillation rate $\delta$, which is determined by using a method that is explained later in this section, the mapper returns a graph that successfully represents the latent geometry of the given networks [Fig.~\ref{fig4}(a--e)].
We use the network distance between nodes in the distilled network as the metric function.
For each connected component in the distilled network, we select a node and use the distances from the other nodes to the selected node as the filter function within the connected component.
We select one of the two nodes that are farthest apart in the connected component of the distilled network.
We then apply the mapper algorithm to the distilled network.

We choose the optimal distillation rate that persistently returns a consistent network structure for a wide range of distillation rates $\delta$.
We use the distances between pairs of nodes to quantify the similarity between the two network structures.
Suppose that $G_1$ and $G_2$ are distilled networks of $G$ with distillation rates $\delta_1$ and $\delta_2>\delta_1$.
The two networks have the same set of nodes, and if a link $\{i,j\}$ is in $G_2$, it is also in $G_1$.
$d_1(i,j)$, which is the distance between two nodes $i$ and $j$ in $G_1$, is no greater than $d_2(i,j)$, the distance between the two nodes in $G_2$.
We quantify the similarity between the two networks $G_1$ and $G_2$ using the Pearson correlation coefficient $r(\delta_1,\delta_2)$ of distances for which $1<d_2<\infty$, which implies  $d_1<\infty$ since $G_1$ is a subgraph of $G_2$.
We exclude the pairs that are directly linked in $G_2$ because they are guaranteed to be linked in $G_1$ and thus do not provide any meaningful information regarding the structure.
We illustrated $r(\delta_1,\delta_2)$ for a Watts–Strogatz small-world network with a ring latent geometry [Fig.~\ref{fig4}(h)].
The plot shows a persistent interval of distillation rate $\delta$, and in this interval, the map yields a correct representation of the latent structure.
We select the parameter $\delta$ that yields the highest consistency, quantified by
\begin{equation}
\chi_{0.025} = \min\{a|r(\delta,\delta+a)>0.975 \,, r(\delta-a,\delta)>0.975\} \,.
\label{consistency}
\end{equation}
This method could potentially be used for parameter selection in the TDA of data points, such as k-nearest neighbor graph construction.

To test the proposed method, we mapped the latent geometries of the networks described in Sec.~\ref{sec:noisy_geometric} and illustrated in Fig.~\ref{fig2} (Networks A--E).
The results are illustrated in Fig.~\ref{fig4}(a--e).
As illustrated in Fig.~\ref{fig4}(f), the Pearson correlation coefficient between the distance in the latent geometry $d(x,y)$ and the distance in the distilled network $d_\delta(x,y)$ is 0.9995 at the optimal distillation rate.
The correlation of the distances between pairs of nodes in the two distilled networks is illustrated in Fig.~\ref{fig4}(g).
There is a persistent interval of the distillation rate $\delta$, in which the distilled network yields a consistent network structure.
The consistency defined in Eq.~\eqref{consistency} shows a clear peak at $\delta=0.31$, and we choose this value as the optimal distillation rate.
In contrast, no persistent interval appears in the $r(\delta_1,\delta_2)$ diagram for the Erd{\H o}s--R\'enyi network, which has no latent geometry.
The same process has been used to determine the optimal distillation rates for other networks.

\section{Topological analysis of empirical networks}
\label{sec:numerical_result}

\begin{figure*}
	\centering
	\includegraphics{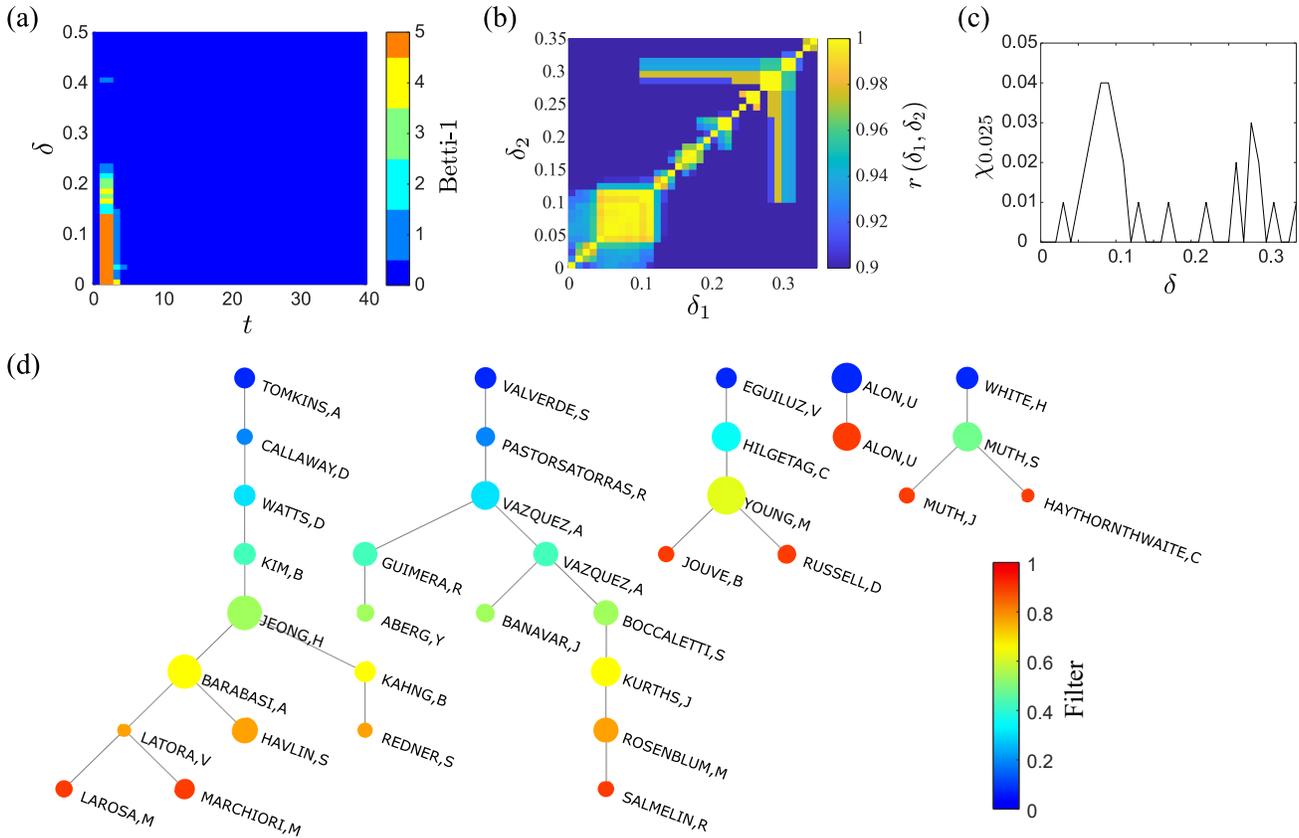}
	\caption{
		(a) Modified persistent homology diagram of the co-authorship structure of researchers in network science.
		No persistent region appears in the diagram, which indicates that there is no loop in the latent geometry.
		(b) Pearson correlation coefficient of distances on distilled networks $r(\delta_1,\delta_2)$.
		A clear persistent interval appears, which indicates that the latent geometry of the network is not completely random.
		(c) The persistent interval measured by $\chi_{0.025}$.
		A clear peak appears at $\delta=0.1$.
		(d) The map of the latent geometry of the co-authorship in network science.
		The size of a vertex in the map is proportional to the number of network nodes in the cluster, and the color encodes the value of the filter function, with red indicating high values and blue indicating low values.
		Five largest connected components in the map are shown.
		The name of the researcher who has the highest degree within the cluster is shown.
		A researcher can simultaneously be in two adjacent clusters because of the overlapping interval of the filter function; therefore, if the researcher has the highest degree within each cluster, the name of the researcher can appear multiple times on the map.
	}
	\label{fig5}
\end{figure*}

To further test the proposed methods, we applied them to a paradigmatic network data set: the co-authorship structure of researchers in network science~\cite{newman2006}.
In this network, there are 1589 scientists from a variety of fields who published an article on the topic of networks, 379 of which fall into the largest connected component.
If two researchers co-authored an article by May 2006, they are connected in the network.
The network is undirected and unweighted.

Figure~\ref{fig5}(a) illustrates the modified persistent homology diagram of the network.
There is no persistent region of nonzero homology in this diagram, and we conclude that there is no loop in the latent geometry of the network.
The Pearson correlation coefficient of the distances in the distilled networks $r(\delta_1, \delta_2)$ is illustrated in Fig.~\ref{fig5}(b).
A clear persistent interval appears around $\delta=0.1$ in the diagram.
This suggests that the latent geometry of the network is not completely random.
The map of the latent geometry is illustrated for the optimal distillation rate $\delta=0.1$ in Fig.~\ref{fig5}(d).
We showed the five largest connected components in the map and displayed the researcher with the highest degree within the cluster as the representative.
Note that a single name can appear multiple times on the map.
This is because the nodes are divided into overlapping sets that correspond to overlapping intervals of the filter function.
Hence, if a researcher that is included in two adjacent clusters has the highest degree within each cluster, the name of the researcher can appear twice on the map.
An edge between the clusters only implies that at least one researcher belongs to both of them and does not necessarily imply a co-authorship of the two representatives of the clusters.

Although there is no previous result on the homology of the latent geometry of the co-authorship structure, one might expect a large loop in the latent geometry of the co-authorship network because the populous region in the globe forms a ring.
However, the modified persistent homology diagram suggests that there is no large-scale loop in the latent geometry of the co-authorship network and rejects such a hypothesis.

\section{Conclusions}
\label{sec:conclusion}

In summary, we developed methods that estimate the latent geometry of complex networks: the modified persistent homology diagram and the map of the latent geometry.
Based on the fact that long-range links have disproportionately higher loads than those of short-range links, we remove links with high loads to obtain a distilled network.
We use two-dimensional filtration for the persistent homology.
A simplicial complex is determined by the distillation rate $\delta$ and radius $t$.
We estimate the homology of the latent geometry by the persistent region in the $\delta$-$t$ plane that yields a constant homology in the modified persistent homology diagram.
This method is completely nonparametric.
We also implemented the mapper to the distilled network to obtain a map of the latent geometry.
Only two parameters must be chosen to map the latent geometry of the network: the length scale of interest and the overlapping percentage of the intervals.
The number of parameters is smaller compared to the standard mapper algorithm, for which the user must choose the length of the interval, overlapping percentage, clustering algorithm and its parameters, and filter function.
We reduce the number of parameters by using a method to determine the optimal distillation rate: we select the distillation rate that yields the most consistent network structure for a wide interval of the parameter.
We use the Pearson correlation coefficient of the distances between pairs of nodes to quantify the consistency of the network structures.

We tested the proposed methods on various synthetic and empirical networks.
For synthetic networks, in which latent geometries are known, the methods successfully estimated the number of loops in the latent geometry and generated the map of the latent geometry.
We also applied the methods on the co-authorship structure of researchers in network science to reveal the latent geometry of the collaboration.
There has been no study on the homology of the latent geometry of the co-authorship structure.
One might expect a large latent loop structure in the co-authorship network because the populous region in the globe forms a ring; however, the modified persistent homology diagram suggests the absence of a loop in the latent geometry and rejects such a hypothesis.
We generated the map of the latent geometry of the co-authorship network for the first time.

Mapping networks to hyperbolic spaces has a wide range of applications.
Such applications include efficient navigation in complex networks such as the Internet ~\cite{boguna2010}, missing link prediction~\cite{wang2016}, and brain mapping~\cite{cacciola2017}.
The latent geometry, or similarity space, assumed for hyperbolic embedding is a ring; therefore, if one tries to embed a network with a line or a double-ring latent geometry on the hyperbolic space, the nodes that are actually distant in the latent geometry can be embedded closely in the hyperbolic space.
The proposed methods cannot estimate the detailed position of the nodes in the latent geometry but can discover the topology of the network.
One can first determine the topology of the latent geometry of a given network and then estimate the position of each node in the latent geometry to perform navigation, link prediction, and mapping on the network.
An interesting work for the future might be the to extend this framework to hypergraphs~
\cite{jhun2019,carletti2020,landry2020,neuhauser2020,dearruda2020,lee2021a,antelmi2021} and simplicial complexes~\cite{courtney2016,iacopini2019,hernandezSerrano2020,matamalas2020,bianconi2020,millan2020,li2022}, which recently attracted considerable interest in the research community~\cite{battiston2020,battiston2021,bianconi2021}.

\begin{acknowledgments}
This research was supported by the NRF, Grant No. NRF-2014R1A3A2069005.
\end{acknowledgments}

\section*{Data Availability Statement}

The data that support the findings of this study are openly available in \url{http://www-personal.umich.edu/~mejn/netdata/}, Ref.~\cite{newman2006}.

\end{document}